\documentclass[preprint,prb,12pt]{revtex4}%
\usepackage{amsfonts}
\usepackage{amsmath}
\usepackage{amssymb}
\usepackage{graphicx}%
\providecommand{\U}[1]{\protect\rule{.1in}{.1in}}

\begin{document}
\preprint{UATP/0904}
\title{Comment on "Comment on: On the reality of residual entropies of glasses and
disordered crystals"\ [J. Chem. Phys. \textbf{129}, 067101 (2008)]}
\author{P. D. Gujrati}
\affiliation{Departments of Physics and Polymer Science, The University of Akron, Akron, OH 44325}

\begin{abstract}
By using very general arguments, we show that the entropy loss conjecture at
the glass transition violates the second law of thermodynamics and must be
rejected. 

\end{abstract}
\date[August 29, 2009]{}
\maketitle

\begin{figure}
[ptb]
\begin{center}
\includegraphics[
trim=0.996427in 6.985081in 2.994169in 1.001365in,
height=2.4163in,
width=3.7689in
]%
{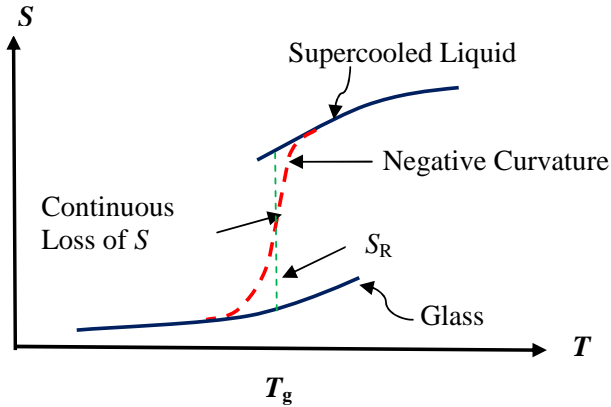}%
\caption{Schematic form of an abrupt entropy drop (dashed green vertical line)
of $S_{\text{R}}$ at the glass transition $T_{\text{g}}$, and a probable
schematic continuous variation (dashed red curve) for a finite system that
avoids the sharp discontinuity$.$ The entropy of the supercooled liquid and of
the glass is shown by blue curves. }%
\label{Fig_EntropyLoss}%
\end{center}
\end{figure}

Gupta and Mauro\cite{Guptaa} in their comment on a recent paper by
Goldstein\cite{Goldstein} explain their entropy loss conjecture (ELC) as
follows: when a glass (GL) is held at a fixed temperature for a period longer
than the experimental time scale $\tau$ used for its preparation, it tends
spontaneously towards the equilibrium state of the supercooled liquid (SCL),
the process known as the structural relaxation. They also claim that the
configurational entropy increases during this relaxation process. This they
say follows from the process being spontaneous and the second law, the law of
increase of entropy: see the figure in their paper.\cite{Guptaa} They also
identify the glass transition as the inverse of relaxation. Hence, they claim
that the configurational entropy $S^{\text{(c)}}$ must decrease during the
glass transition. This seems to be their justification of ELC. The situation
with a discontinuous entropy loss is schematically shown in Fig.
\ref{Fig_EntropyLoss}, where the blue curves show the entropies of SCL and the
GL, and the dashed green vertical jump shows the discontinuity of magnitude
$S_{\text{R}}$ at the glass transition at $T_{\text{g}}.$ For a finite system
such as those investigated in simulations, this discontinuity most probably
will be replaced by a continuous piece shown by the dashed red curve.

\begin{figure}
[ptb]
\begin{center}
\includegraphics[
trim=1.497492in 5.484629in 1.897530in 1.302520in,
height=3.1107in,
width=3.8303in
]%
{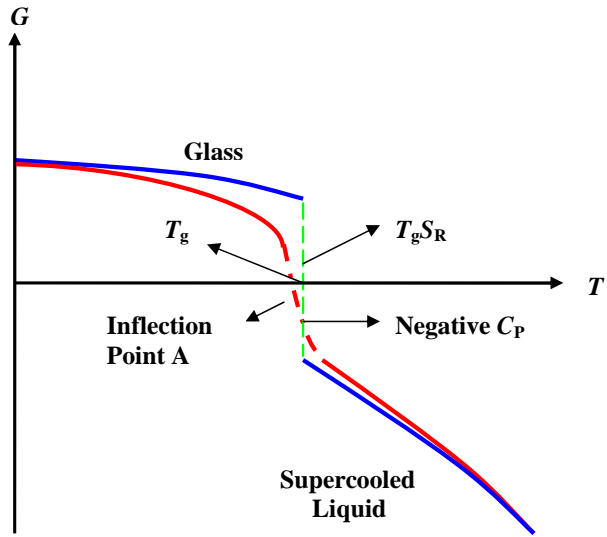}%
\caption{Schematic form of the concave free energy $G$ as a function of the
temperature $T$. The blue curve represents the Gibbs free energy for a
macroscopically large system; the green discontinuity $T_{\text{g}}%
S_{\text{R}}$ at $T_{\text{g}}$ shown by the broken green portion corresponds
to the residual entropy loss $S_{\text{R}}$. (This discontinuity will be
absent if there were no entropy loss.) For finite systems, this discontinuity
will be most probably replaced by a continuous curve shown by red, which can
be conveniently divided into three different pieces: the two solid red pieces
connected by a dot-dashed piece with an inflection point A. This continuous
curve has a region of negative heat capacity, making it unphysical.}%
\label{Fig_Gibbs_Free_Energy}%
\end{center}
\end{figure}

It is agreed by all, and this is also supported by experiments, that the
enthalpy $H$ and the volume $V$ show no discontinuity at the glass transition.
Consequently, the entropy discontinuity gives rise to a discontinuous jump in
the Gibbs free energy $G=H-TS$ in the amount of $T_{\text{g}}S_{\text{R}}$, as
shown by the blue curves and the green discontinuity in Fig.
\ref{Fig_Gibbs_Free_Energy}. The red curve shows the Gibbs free energy when
the entropy loss in Fig. \ref{Fig_EntropyLoss} is continuous. We will discuss
the continuous behavior later. For the moment, we are interested in the
discontinuity behavior of ELC discussed by Gupta and Mauro.\cite{Guptaa}

Goldstein\cite{Goldstein} had used some plausible assumptions and Henry's law
to argue that ELC would lead to the violation of the second law. Gupta and
Mauro\cite{Guptaa} countered by claiming that the glass transition is not a
spontaneous process, so the discontinuous drop in its entropy is not a
violation of the second law. Which claim is right? How should we resolve this
impasse? Any attempt to resolve the issue must be based on some general
arguments which not only bypasses the limitations of the arguments used by
Goldstein, but also bypasses any assumption of whether the glass transition is
spontaneous or not. In this comment, we establish on very general grounds that
the above discontinuity in the Gibbs free energy violates the second law. As
this discontinuity is a consequence of ELC, ELC itself cannot be valid.%
\qquad\qquad

The glass is a system far from equilibrium so one may not apply equilibrium
statistical mechanics or equilibrium thermodynamics to investigate its
properties, which vary with time. One may resort to apply non-equilibrium
thermodynamics, not a well-developed field at present, to study glasses and
their relaxation in time. To avoid discussing whether the glass transition is
a spontaneous process or not, we consider an isolated system $\Sigma_{0}$
composed of the system $\Sigma$ of interest (GL/SCL) in a medium denoted by
$\widetilde{\Sigma}$. We also do not commit as to how the entropy is
calculated in order to avoid the contentious debate about which formalism is
the correct one, except to demand that the formalism is equally applicable to
SCL and GL, and that it satisfies the second law. We then follow the
consequences of the second law applied to the isolated system. We will
consider a single component system, which is sufficient for our purpose.
According to this law, the entropy $S_{0}(t)$ of an isolated system
$\Sigma_{0}$ can never decrease in time.\cite{Landau} In general, $S_{0}$ also
depends on the number of particles $N_{0}$, energy $E_{0}$, and volume $V_{0}$
of $\Sigma_{0}$. Thus, $S_{0}(t)$ used above should be really written as
$S_{0}(E_{0},V_{0},N_{0},t)$. However, as the extensive quantities remain
constant in time there is no harm in using the compact form $S_{0}(t)$ during
approach to equilibrium.\ The energy, volume and the number of particles of
$\Sigma$\ are denoted by $E$, $V$, and $N,$ respectively, while that of the
medium $\widetilde{\Sigma}$ by $\widetilde{E}$, $\widetilde{V}$, and
$\widetilde{N}.$ Obviously,%
\[
E_{0}=E+\widetilde{E},\ \ V_{0}=V+\widetilde{V},\ \ N_{0}=N+\widetilde{N}.
\]
We will assume that $N$ of the system is also fixed, which means that
$\widetilde{N}$ is also fixed. However, the energy and volume of the system
may change with $t$.

When the isolated system is in equilibrium, its entropy $S_{0}(E_{0}%
,V_{0},N_{0},t)$ has reached its maximum and no longer has any explicit
time-dependence so that it can be simply written as $S_{0}(E_{0},V_{0},N_{0})$
or $S_{0}$. In this case, different parts of $\Sigma_{0}$\ have the same
temperature $T=(\partial E_{0}/\partial S_{0})$ and pressure $P=T(\partial
V_{0}/\partial S_{0})$.\ Otherwise, the entropy $S_{0}(t)$ \emph{continuously
increases} and the isolated system is said to be not in equilibrium. The
medium is considered to be very large compared to $\Sigma,$ so that its
temperature, pressure, etc. are not affected by the system. We assume
$\widetilde{\Sigma}$ to be in internal equilibrium (its different parts have
the same temperature and pressure, but $\widetilde{\Sigma}$ and $\Sigma$ may
not be in equilibrium with each other). Thus, its entropy no longer has an
explicit time dependence, but has implicit $t$-dependence through the
$t$-dependence of $\widetilde{E}$, and $\widetilde{V}$. Below the glass
transition at $T_{\text{g}}$, GL ($\Sigma$) will relax so as to come to
equilibrium with the medium if we wait longer than $\tau$.

The entropy $S_{0}$ of the isolated system can be written as the sum of the
entropies $S$ of the system and $\widetilde{S}$ of the medium:%
\[
S_{0}(E_{0},V_{0},N_{0},t)=S(E,V,N,t)+\widetilde{S}(\widetilde{E}%
,\widetilde{V},\widetilde{N});
\]
there is no explicit $t$-dependence in $\widetilde{S}(\widetilde{E}%
,\widetilde{V},\widetilde{N})$ due to internal equilibrium. The correction to
this entropy due to the weak stochastic interactions between the system and
the medium has been neglected, which is a common practice.\cite{Landau} We
expand $S_{0}$ in terms of the small quantities of the system\cite{Landau}%
\[
\widetilde{S}(\widetilde{E},\widetilde{V},\widetilde{N})\simeq\widetilde
{S}(E_{0},V_{0},\widetilde{N})-\left.  \left(  \frac{\partial\widetilde{S}%
}{\partial\widetilde{E}}\right)  \right\vert _{E_{0}}E(t)-\left.  \left(
\frac{\partial\widetilde{S}}{\partial\widetilde{V}}\right)  \right\vert
_{V_{0}}V(t).
\]
It follows from the internal equilibrium of $\widetilde{\Sigma}$ that%
\begin{equation}
\left.  \left(  \frac{\partial\widetilde{S}}{\partial\widetilde{E}}\right)
\right\vert _{E_{0}}=\frac{1}{T},\ \ \left.  \left(  \frac{\partial
\widetilde{S}}{\partial\widetilde{V}}\right)  \right\vert _{V_{0}}=\frac{P}%
{T}, \label{Medium_Relations}%
\end{equation}
and $\widetilde{S}\equiv\widetilde{S}(E_{0},V_{0},\widetilde{N}),$ which is a
constant, is independent of the system. Thus,%
\begin{equation}
S_{0}(t)-\widetilde{S}\simeq S(E,V,N,t)-E(t)/T-PV(t)/T.\nonumber
\end{equation}

Let us introduce
\begin{equation}
G(t)\equiv H(t)-TS(t),\ H(t)\equiv E(t)+PV(t), \label{Free_Energies}%
\end{equation}
the time-dependent Gibbs\ free energy and enthalpy of the system $\Sigma$ with
the medium $\widetilde{\Sigma}$ at fixed $T$ and $P$. In terms of these
functions, we have
\begin{equation}
S_{0}(t)-\widetilde{S}=S(t)-H/T=-G/T.
\label{Gibbs_Free_Energy_Entropy_Relation}%
\end{equation}
In deriving the above equation (\ref{Gibbs_Free_Energy_Entropy_Relation}), no
assumption about the system $\Sigma$ has been made. In particular, we have not
assumed any particular aspect of its non-equilibrium nature. We have also not
assumed whether the glass transition is spontaneous or not. The identification
of $S_{0}(t)-\widetilde{S}$ with the Gibbs free energy of $\Sigma$ is
generally valid under the assumption of the medium being large compared to
$\Sigma$, which can be satisfied as well as we wish. Now, if $G$ underwent a
discontinuous jump upwards as SCL turns into GL at the glass transition in a
cooling experiment, see Fig. \ref{Fig_Gibbs_Free_Energy}, it will lower the
entropy of the isolated system by $S_{\text{R}}$. But, this is inconsistent
with the second law! Therefore, there cannot be an entropy loss in the system
as SCL turns into GL at the glass transition. This proves our above claim.

The above violation of the second law occurs because of the discontinuity. The
violation will not occur if the jump is replaced by a continuous patch as
shown in Figs. \ref{Fig_EntropyLoss} and \ref{Fig_Gibbs_Free_Energy}. Now, the
Gibbs free energy must be a concave function of the temperature so that the
heat capacity remains non-negative, as the blue curves are in Fig.
\ref{Fig_Gibbs_Free_Energy}. The free energy must always curve downwards for
$C_{\text{P}}$ to remain non-negative. The red curve contains three different
pieces:\ the two solid pieces in Fig. \ref{Fig_Gibbs_Free_Energy} on either
side of the dot-dashed piece in the middle containing an inflection point A
where the curvature changes its sign. The solid pieces are concave as required
by stability. But it is also clear that any attempt to smoothly connect the
two solid pieces by the dot-dashed piece near $T_{\text{g}}$ must result in
the middle piece, which cannot remain concave everywhere. It must have a
convex piece at higher temperatures, which will then result in a region of
\emph{negative heat capacity} as shown by the arrow, and which will make the
system \emph{unphysical}. Recently, Mauro et al \cite{Mauro} have reported
model computation results for a finite system\ of selenium by using several
cooling rates. They observe a continuous drop in the configurational entropy.
Unfortunately they have not carried out any finite size analysis, so it is not
possible to know what happens to the continuous drop in the limit of a
macroscopically large system. They have also not reported the heat capacity or
the entropy so no comment can be made regarding their behavior. We will have
to wait for their answer.

\end{document}